\documentclass[twocolumn]{revtex4}
\usepackage [cp1251]{inputenc}
\usepackage [english]{babel}
\usepackage{amsmath}
\usepackage{bm}

\usepackage{color}
\usepackage{epsfig}
\usepackage[colorlinks=true,citecolor=blue,linkcolor=blue, urlcolor=cyan]{hyperref} 
\usepackage{orcidlink}

\newcommand{\nix}[1]{}
\usepackage{bbold}

\newcommand{\addMisha}[1]{\textcolor{black}{#1}}

\usepackage{soul}

\begin{document}

\title{Photocurrents induced by ${\bm k}$-linear terms in semiconductors and semimetals}
\author{M.~M.~Glazov\orcidlink{0000-0003-4462-0749}, E.~L.~Ivchenko\orcidlink{0000-0001-7414-462X}}
\affiliation{Ioffe Institute, 194021, St.-Petersburg, Russia}

\begin{abstract}
We develop a six-band $\bm{k} \cdot \bm{p}$ model to describe the electronic structure and optical response of chiral multifold semimetals, such as RhSi. By means of invariants method we construct the effective Hamiltonian describing the states near the $\Gamma$-point of the Brillouin zone where the spin-orbit coupling and $\bm{k}$-linear Rashba terms, which are crucial for circular photogalvanic effect, are taken into account. The model is parameterized using tight-binding calculations. We compute the interband absorption spectrum, showing a linear-in-frequency dependence at low energies and a resonant feature near the spin-orbit splitting energy. Furthermore, we calculate the circular photogalvanic effect. In agreement with previous works the current generation rate at low frequencies exhibits a quantized low-frequency response, governed by the universal value $|\mathcal{C}| = 4$ for the effective topological charge.  Our results provide an analytical framework for understanding the role of Rashba coupling and topology in the optoelectronic properties of multifold chiral semimetals.
\end{abstract}
  \maketitle

\emph{Introduction.} Rashba and Sheka were the first who attracted the attention to an existence of spin-dependent terms linear in the electron wave vector ${\bm k}$ in the $\Gamma_7$ conduction and valence bands of wurtzite crystals \cite{Rashba-1959-II}. This work became the starting point for studies of consequences of the $\bm k$-linear contribution ${\cal H}^{(1)}({\bm k}) =\beta_{ij} \sigma_i k_j$ to the free-charge carrier effective Hamiltonian, where $\sigma_i$ are the Pauli spin matrices and $\beta_{ij}$ are band parameters which form a second-rank pseudotensor. This tensor couples the orbital and spin dynamics, i.e., components of the polar vector ${\bm k}$ with the axial-vector components $\sigma_i$, and is nonzero for point groups that allow optical activity or gyrotropy \cite{LandauLifshitz}. In fact, the spin-dependent linear-in-${\bm k}$ terms are an electronic analog of mechanical systems where a rotatory motion is mixed with a linear one like it occurs in the propeller and wheel effects. 

The first consequence of the spin-orbit term ${\cal H}^{(1)}({\bm k})$ was the combined resonance predicted by Rashba in Ref.~\cite{Rashba1960} which is an excitation of spin transitions by the electric component of the electromagnetic field. Another example is the resonant optical manifestation of terms linear in the exciton center-of-mass wave vector ${\bm K}$, namely, anomalous features in the 
reflectivity spectra near the B-exciton resonance frequency \cite{Mahan} and the $s$-$p$ polarization conversion under oblique light reflection \cite{Sel'kin,ELIvchenko}. The ${\bm k}$-linear terms in the free-charge carrier Hamiltonian play the main role in the circular photogalvanic effect (CPGE) predicted in Ref.~\cite{IvchenkoPikus} and observed first by Asnin et al. \cite{Asnin}, as well as in recently proposed circular Raman photogalvanic effect~\cite{gg22}. The CPGE has been studied not only in semiconductors and semiconductor nanostructures, see Refs.~\cite{Ivchenko,golub_ganichev_BIASIA} for review, but also in three-dimensional Weyl semimetals \cite{Moore,Weyl-review,Hasan2018,Golub-Weyl-1,Golub-Weyl-2}. It has been shown that in the absence of inversion and mirror symmetries, the circular photocurrent in such systems is quantized in units of the fundamental physical constants~\cite{Moore}.

In recent years, along with Dirac and Weyl semimetals, a new class of topological semimetals with chiral symmetry and multiple degeneracy points in the Brillouin zone -- multifold semimetals -- has been studied, see reviews \cite{Review2021,Bernevig Review}. Among them, a number of non-magnetic semimetals with spatial symmetry P2$_1$3 (the space group N = 198, the chiral point group T) stands out: a family of binary compounds MgPt, RhSi, RhSn, PdGa, PtGa, CoSi 
\cite{Review2021, Bernevig Review, Bernevig2016, Hasan2017, RSchW2017,Burkov2018,Burkov2019,PdGa 2021,Stishov2023}, as well as ternary materials PdBiSe, SrGePt, CaSiPt, \ldots  \cite{SrGePt 2023}. This symmetry allows for topological points (or nodes) of four- and six-fold degeneracy and the corresponding Fermi arches of surface states \cite{RhSi 2021,PtGa-arc}. As in other gyrotropic crystals, the {CPGE can} be and is indeed observed in semimetals with the point symmetry T \cite{Review2021, Hasan2017, Sun2020, Grushin2018, Grushin2019, RhSi 2021, Mele2021,pronin-book}. The theoretical description of the energy spectrum and CPGE in these materials has so far been carried out using microscopic numerical calculations. In this paper, we will use the effective Hamiltonian method for such a description, which allows us to obtain analytical results and underline the importance of the $\bm k$-linear, Rashba, terms in the effect.

As an example of the efficiency of the method, we consider the electron states in the vicinity of the center of the Brillouin zone, $\Gamma$. It is assumed that without taking into account the spin and spin-orbit interaction, the energy state $E_0$ in the $\Gamma$ point is threefold degenerate and effectively characterized by the angular momentum $L = 1$, the representation $\Gamma_4$ of the T group. With allowance for the spin, the degeneracy increases to six. Taking into account the spin-orbit interaction, the degeneracy is partially lifted, resulting into the quartet $\Gamma_6 + \Gamma_7$ and the doublet $\Gamma_5$, they can be assigned to the angular momenta $J= 3/2$ and $J= 1/2$, respectively. We expand the effective 6$\times$6 Hamiltonian ${\cal H}({\bm k})$ in powers of the electron wave vector ${\bm k}$, preserving the first- and second-order terms, and find linearly independent expansion coefficients using the invariant method \cite{BirPikus, Book1997, Winkler}. By comparing the results of calculations of the electron band structure using atomistic modeling and the effective Hamiltonian method, we can parameterize the ${\bm k}\cdot {\bm p}$ Hamiltonian ${\cal H}({\bm k})$ and calculate spectra of the optical absorption and circular photocurrent. \addMisha{The application of this approach is especially effective for the silicides RhSi and CoSi, for which the $\Gamma_6 + \Gamma_7$ and $\Gamma_5$ bands are adjacent to the Fermi energy  while the role of other bands is diminished.}

\emph{$\bm k\cdot \bm p$-model.} Let us denote by $X,Y,Z$ the orbital functions transforming according to the representation $\Gamma_4$ as coordinates $x \parallel [100]$, $y \parallel [010]$, $z \parallel [001]$. For spin states $\pm 1/2$, we introduce two basic spinors $\alpha_{+\frac{1}{2}}\equiv \alpha {=\ \uparrow}$ and $\alpha_{-\frac{1}{2}} \equiv \beta {=\ \downarrow}$. The six products $\alpha_s R_{\eta}$ ($s= \pm 1/2$, $\eta = x,y,z$), $R_x = X, R_y = Y, R_z = Z,$
form the basis of the representation $\Gamma_4 \times \Gamma_5$ of group T, which decomposes into the irreducible spinor representations $\Gamma_5 + \Gamma_6 + \Gamma_7$. We expand the basis functions of these irreducible representations $\Psi_j$ ($j=1,2,\dots,6$) in terms of the functions $\alpha_s R_{\eta}$:
\begin{equation} \label{PsiaR}
\Psi_j = \sum\limits_{s \eta}C^{(j)}_{s, \eta} \alpha_s R_{\eta}.
\end{equation}
For the representation $\Gamma_5$, this expansion can be chosen in the form of basis functions for angular momentum $J=1/2$:
\begin{eqnarray} \label{basis12}
\left \vert \Gamma_{5},+1/2 \right> &=& - \frac{1}{\sqrt{3}} [\alpha Z + \beta (X + {\rm i} Y) ], \\ 
\left \vert \Gamma_{5},-1/2 \right> &=& \frac{1}{\sqrt{3}} [ \beta Z - \alpha (X - {\rm i} Y) ]. \nonumber
\end{eqnarray}
The basis functions of the two-dimensional representations $\Gamma_6$ and $\Gamma_7$ can be constructed using tables of standard basis functions for the point group T \cite{Koster}. However, for convenience in writing the effective electron Hamiltonian $\mathcal{H}(\bm{k})$, we will use functions that are linear combinations of the basis functions of representations $\Gamma_6$ and $\Gamma_7$ and can be conveniently assigned to the angular momentum $J=3/2$ with projections $\pm 1/2, \pm 3/2$, namely:
\begin{eqnarray} \label{basis32}
\left \vert \Gamma_6 + \Gamma_7,+3/2 \right> &=& - \alpha \frac{X + {\rm i} Y}{\sqrt{2}}, \\
\left \vert \Gamma_6 + \Gamma_7,+1/2 \right> &=& \sqrt{\frac{2}{3}} \alpha Z - \beta \frac{X + {\rm i} Y}{\sqrt{6}}, \nonumber \\ 
\left \vert \Gamma_6 + \Gamma_7,-1/2 \right> &=& \sqrt{\frac{2}{3}} \beta Z + \alpha \frac{X - {\rm i} Y}{\sqrt{6}}, \nonumber \\ 
\left \vert \Gamma_6 + \Gamma_7,-3/2 \right> &=& \beta \frac{X - {\rm i} Y}{\sqrt{2}}. \nonumber
\end{eqnarray}
Such a basis is used in the books \cite{Winkler,Ivchenko} for constructing the Luttinger Hamiltonian for the $\Gamma_8$ band in semiconductor crystals of the T$_d$ symmetry with zinc-blende structure.

\begin{figure*}[bt]
\includegraphics[width=0.9\textwidth]{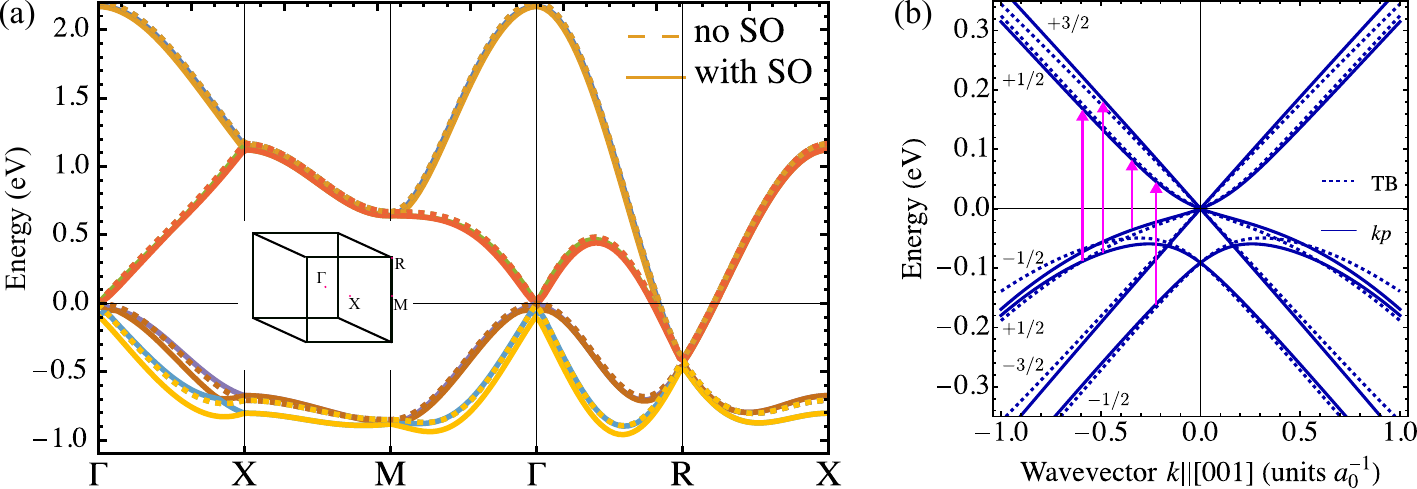}
\caption{(a) Electron dispersion in RhSi found using the tight-binding method according to Ref.~\cite{Hasan2017}. The used tight-binding parameters in eV~\cite{Hasan2017} are $v_1 = 0.55$, $v_2 = 0.16$, $v_p =-0.76$, $v_{r1} = 0$, $v_{r2} =-0.03$, $v_{r3} = 0.01$, $v_{s1} =-0.04$,  $v_{s2} = v_{s3} = 0$. Dashed and solid lines show the calculation neglecting and including spin-orbit interaction. The inset shows the scheme of the first Brillouin zone with high-symmetry points labeled as $\Gamma$, X, M, and R. (b) Spectrum near the $\Gamma$ point calculated with spin-orbit interaction in the tight-binding method (dotted) and in the $\bm{k} \cdot \bm{p}$ model (solid curves). Arrows denote allowed optical transitions in $\sigma^+$ polarization (in the isotropic model). The parameters of the $\bm{k} \cdot \bm{p}$ model obtained by fitting the tight-binding calculation are given in Table~\ref{tab:kp}.}\label{fig:spec}
\end{figure*}

Here we present an expression for the effective Hamiltonian $\mathcal{H}(\bm{k})$ in the $\Gamma_{6} + \Gamma_{7} + \Gamma_5$ bands, obtained by neglecting spin-orbit mixing with states from other (remote) bands. Using the method of invariants, the operator $\mathcal{H}(\bm{k})$ can be written as
\begin{eqnarray} \label{H2}
&&\mathcal{H} = \ell \hspace{0.7 mm}\bm{k}\cdot \bm{I} + \frac{\Delta}{3} \bm{\sigma}\cdot \bm{I} - \frac{\Delta}{3} \\ && \hspace{0.3 cm} {+\ L k^2} - (L-M) \sum\limits_i I_i^2 k_i^2 - N \sum\limits_{i' \neq i} \{ I_i I_{i'} \}_s k_i k_{i'}, \nonumber
\end{eqnarray}
where $\bm{\sigma}$ is a pseudovector whose components are the Pauli spin matrices $\sigma_x, \sigma_y, \sigma_z$; $I_i$ are the angular momentum matrices $L=1$ in the basis of functions transforming as coordinates $x,y,z$:
\begin{equation} \label{I}
I_x = \left[ \begin{array}{ccc} 0&0&0\\ 0&0&-{\rm i}\\ 0&{\rm i}&0 \end{array} \right],~I_y = \left[ \begin{array}{ccc} 0&0&{\rm i}\\ 0&0&0\\ -{\rm i} &0&0 \end{array} \right],~I_z = \left[ \begin{array}{ccc} 0&-{\rm i}&0\\ {\rm i}&0&0\\ 0&0&0 \end{array} \right].\nonumber
\end{equation}
In constructing the Hamiltonian (\ref{H2}), it is taken into account that the matrices for the linear and quadratic terms in $\bm{k}$ must be odd and even under time reversal, respectively.

It is instructive to introduce the 6 basic functions $\Psi_j$ ($j=1$-6) in the following order: the four functions $\Psi_j$ with $j=1$-4 in the expansion (\ref{PsiaR}) correspond to states $\left\vert \Gamma_6 + \Gamma_7, m \right>$ with $m = 3/2, 1/2, -1/2, -3/2$, while the functions $\Psi_5, \Psi_6$ correspond to states $\left\vert \Gamma_5, m' \right>$ with $m' = 1/2, -1/2$. 
In this basis, the effective Hamiltonian takes the form
\begin{eqnarray} \label{H}
&\mathcal{H} = \mathcal{H}^{(0)} + \hbar v_0 \hspace{0.5mm}\bm{k}\cdot \bm{\mathcal{J}} \\ 
& + \left(A + \frac54 B\right) k^2 - B \sum\limits_i \mathcal{J}_i^2 k_i^2 - \frac{D}{\sqrt{3}} \sum\limits_{i' \neq i} \{ \mathcal{J}_i \mathcal{J}_{i'} \}_s k_i k_{i'}. \nonumber
\end{eqnarray}
Here, the energy is measured from the energy of the $\Gamma_6 + \Gamma_7$ states, the matrix $\mathcal{H}^{(0)}$ is a diagonal $6\times 6$ matrix with only two non-zero elements being $-\Delta$, where $\Delta$ is the spin-orbit splitting at the $\Gamma$-point, 
\[
 A = \frac{L + 2 M}{3}\:,\:  B = \frac{L -M}{3}\:,\: N = \sqrt{3}D\:,
 \]
$\hbar v_0 = \frac23 \ell$, and the six-dimensional matrices $\mathcal{J}_i$ are related to the matrices $I_i$ by (see the Supplementary Materials (SM) ~\cite{SM} for their explicit form):
\begin{equation} \label{JI}
\mathcal{J}_{i; j'j} = \frac32 \sum_{s,\eta,\eta'} C^{(j')*}_{s,\eta'} I_{i; \eta' \eta} C^{(j)}_{s,\eta}.
\end{equation}
When spin-orbit mixing with remote bands is taken into account, the structure of the matrix $\mathcal{H}^{(1)}(\bm{k})$ becomes more complicated. Its general form, obtained by the method of invariants, is given in the SM~\cite{SM}. Sometimes when discussing electronic quasiparticles with $J=3/2$, an analogy is made with elementary particles with spin 3/2 and the term ``Rarita-Schwinger-Weyl fermions'' is used~\cite{RSchW2017,PtGa-arc,RSchW2016,RShW2020,Haidukov}. For $\hbar v_0 k$ or $Bk^2, Dk^2$ comparable with $\Delta$, instead of $4\times4$ Hamiltonian one should consider a six-band model with Hamiltonian \eqref{H2} and this analogy loses meaning. Such electrons can be called $\left(1,\frac12\right)$-fermions. In contrast to the quasiparticles in crystals, the spin-3/2 relativistic particles obey Lorentz invariance and do not have any fine structure in the absence of external fields, see \S31 ``The wave equation for a particle with spin $3/2$'' in Ref.~\cite{ll4_eng}.

Figure~\ref{fig:spec} shows the electron dispersion in the RhSi crystal. Panel (a) demonstrates the energy bands obtained within the tight-binding method proposed in~\cite{Hasan2017} (see also~\cite{Grushin2018}). This model agrees well with density functional theory and adequately describes the energy spectrum over a wide energy range. To illustrate the role of spin-orbit coupling, the dashed curves in Fig.~\ref{fig:spec}(a) show the results of calculations neglecting spin-orbit interaction. It can be seen that on energy scales of $\sim$1 eV, spin-orbit coupling does not play a decisive role, but it determines the fine structure of the spectrum near the $\Gamma$-point of the Brillouin zone that interests us, Fig.~\ref{fig:spec}(b).

To parameterize the $\bm{k} \cdot \bm{p}$ model, we expanded the tight-binding Hamiltonian given in Sec.~III of SM near $\bm{k} = 0$ up to quadratic terms in $\bm k$. Comparing the calculated spectrum with that of the effective Hamiltonian~\eqref{H} allowed us to determine the parameters of the $\bm{k} \cdot \bm{p}$ model, see Table~\ref{tab:kp}. The best agreement within the calculation error is for the isotropic approximation $D = \sqrt{3}B$, or $N = L-M$ in the notation of Eq.~\eqref{H2}. In this approximation the electronic states are characterized by a definite angular moment component $\pm 1/2, \pm 3/2$ onto the ${\bm k}$ vector and their energy $E_n({\bm k})$ is independent of the direction of ${\bm k}$. It is also interesting to note that within this parameterization, accounting for spin-orbit mixing with remote bands is not required: all linear $\bm{k} \cdot \bm{p}$ terms are determined by the non-relativistic constant $\hbar v_0$. Figure~\ref{fig:spec}(b) shows a comparison of the $\bm{k} \cdot \bm{p}$ calculation (solid lines) with the tight-binding calculation for $\bm{k} \parallel [001]$. Analytical expressions for the energy spectrum are given in the SM~\cite{SM}.

\begin{table}[b]
\caption{Parameterization of the $\bm{k} \cdot \bm{p}$ model based on the tight-binding method.  The lattice constant $a_0 = 4.67$~\AA~\cite{Grushin2018,Geller1954}, the values of tight-binding parameters $v_1, v_2$ and $v_p$ are presented in caption to Fig.~\ref{fig:spec}. The cumbersome expression for $\Delta$ in terms of parameters $v_i$ is given in the SM~\cite{SM}.} \label{tab:kp}
 \begin{center}
\begin{tabular}{l||l} \hline 
Parameter & Value \\ \hline \hline
$\Delta$ & $91$~meV\\ \hline
$\ell$ & $\frac{v_p a_0}{2}$\\ \hline
$L$ & $\frac{v_2a_0^2}{2} - \frac{v_p^2 a_0^2}{16 v_1}$\\ \hline
$M$ & $\frac{v_1 a_0^2}{4}-\frac{v_2a_0^2}{2} - \frac{v_p^2 a_0^4}{16 v_1}$\\ \hline
$N$ & $L-M$ \\ \hline
\end{tabular}
\end{center}
\end{table}

\begin{figure}[h!]
\includegraphics[width=0.9\linewidth]{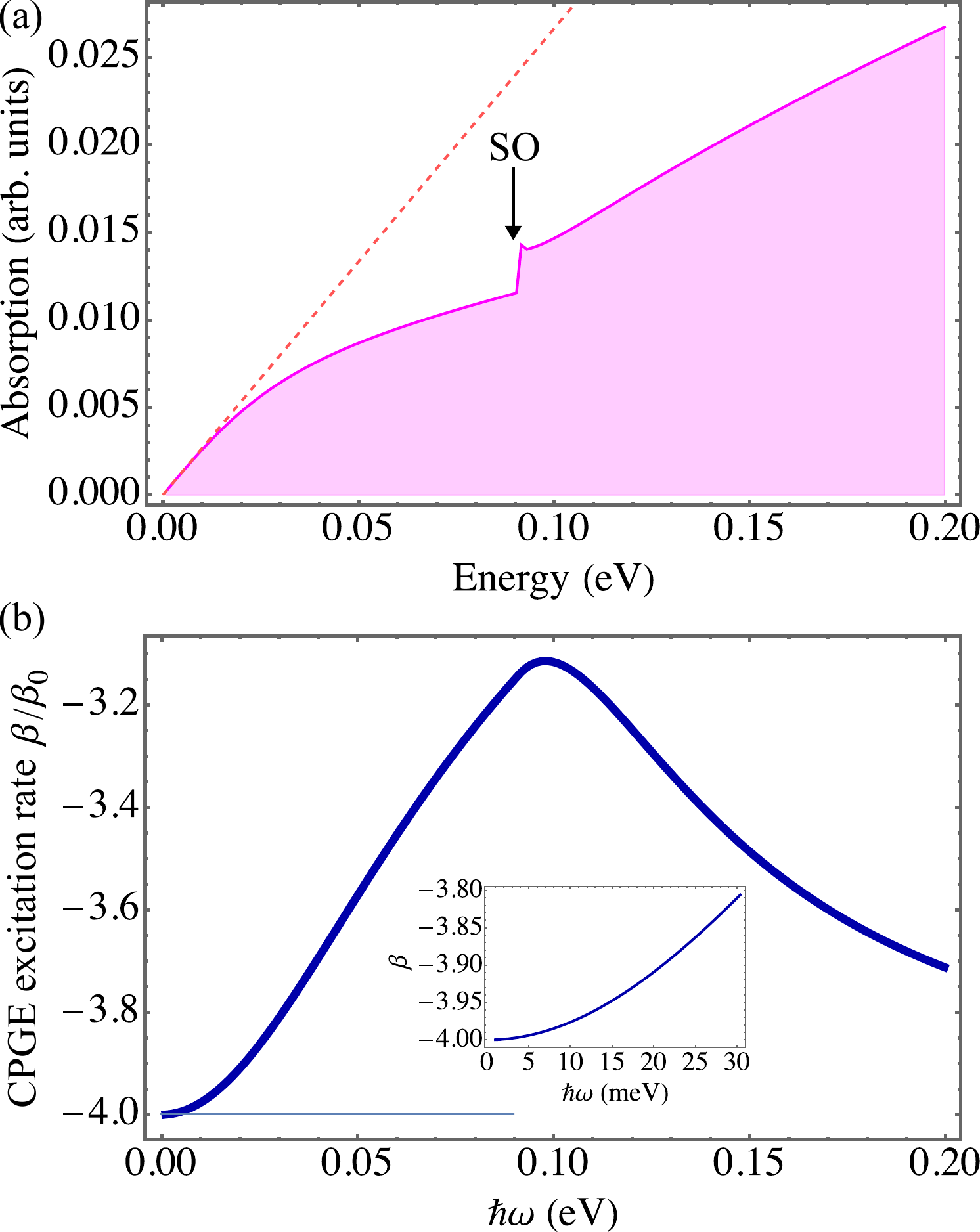}
\caption{(a) Absorption spectrum for interband optical transitions calculated within the $\bm{k} \cdot \bm{p}$ model for the Fermi energy $E_F=0$. Dashed line shows the linear in $\omega$ extrapolation. (b) The CPGE excitation rate coefficient $\beta$ related to the quantized current constant. Thin horizontal line shows the universal value $-4$, see Eq.~\eqref{universal} and details in text. Inset shows a zoom-in at low frequencies. }\label{fig:cpge}
\end{figure}

\emph{Interband absorption near the $\Gamma$-point of the Brillouin zone.}
Let us now calculate the absorption spectrum of RhSi crystal for interband transitions. The $6\times6$ velocity operator is written in the standard form:
\begin{equation}
\label{velocity}
\hat{\bm v}({\bm k}) = \frac{1}{\hbar} \frac{\partial \mathcal{H}({\bm k})}{\partial {\bm k}}.
\end{equation}
The matrix elements of the velocity operator are expressed through the Hamiltonian's eigencolumns as
\begin{equation}
{\bm v}_{n'n}({\bm k}) = \hat{C}^{\dag}_{n'}({\bm k}) \hat{\bm v}({\bm k}) \hat{C}_n({\bm k}),
\end{equation}
Accordingly, the contribution to the absorption coefficient related to the transition $n \to n'$ reads
\begin{multline}
\label{Knum}
K_{n'n} = \frac{4\pi^2e^2}{\omega c n_{\omega}V}\sum\limits_{\bm k} | {\bm e}\cdot {\bm v}_{n'n}({\bm k})|^2 \\
\times \left( f_{n k} - f_{n'k}\right) \delta \left[ E_{n'}(\bm k) - E_n(\bm k)- \hbar \omega \right],
\end{multline}
where $\bm e$ is the unit polarization vector of light, $V$ is the sample volume, $f_{nk}$ is the electron distribution function, and $E_n(\bm k)$ is the dispersion of the band~$n$.

Figure~\ref{fig:cpge}(a) shows the absorption spectrum of RhSi due to interband transitions calculated after $K(\hbar\omega) = \sum_{n,n'} K_{n',n}$, where the coefficients $K_{n',n}$ are determined by Eq.~\eqref{Knum}. At small frequencies $K\propto \hbar\omega$ as expected for the bulk systems with $k$-linear spectra. The feature at $\hbar\omega \approx 90$~meV is related to the onset of the absorption due to an inverse square-root singularity in the reduced density of states for the upper and lower bands with the angular-momentum component $\langle  \bm{\mathcal{J}} \cdot {\bm k} \rangle/k = 1/2$. 

\emph{Circular photocurrent in the $\bm{k} \cdot \bm{p}$ model.} Let us now proceed to the main results of our work, namely, the theory of the circular photogalvanic effect (CPGE). Following Ref. \cite{Moore} we calculate the CPGE injection current defined by
\begin{equation} \label{beta}
\frac{d {\bm j}}{dt} = {\rm i}  [{\bm e} \times {\bm e}^*] \beta E^2_0\:.
\end{equation}
Here, as before, $\bm e$ is the unit polarization vector of light, $E_0$ is the field amplitude inside the crystal, the cross product $\mathrm i[{\bm e} \times {\bm e}^*]$ can be expressed through the degree of circular polarization $P_{\rm circ}$ and the light wave vector $\bm q$ as $P_{\rm circ} \bm q/q$. The microscopical parameter $\beta$ is  a sum of partial contributions
\begin{eqnarray}
\label{gamma:nn'}
&&\beta_{n'n} = \frac{ 2\pi} {\hbar} \frac{e^3}{\omega^2 V}\sum\limits_{\bm k}  (v_{n'{\bm k};z} - v_{n{\bm k};z}) |{\bm e}_{\sigma_+}{\bm v}_{n'n}({\bm k})|^2 \nonumber \\ && \hspace{0.8 cm} \times  \left( f_{n \bm k} - f_{n'\bm k}\right) \delta \left[ E_{n'}(\bm k) - E_n(\bm k)- \hbar \omega \right],
\end{eqnarray}
where the $z$-component of group velocity is [cf.~\eqref{velocity}] $v_{n{\bm k};z} = \hbar^{-1} {\partial E_n(\bm k)}/{\partial k_z}$. The unit vector of the $\sigma_+$ circularly polarized light propagating along the $z$-axis is ${\bm e}_{\sigma_+} = (\hat{x} + {\rm i} \hat{y})/\sqrt{2}$, where $\hat{x}, \hat{y}$ are the unit vectors along the corresponding axes.

Figure~\ref{fig:cpge}b presents the results of calculating the circular photocurrent excitation spectrum versus the photon energy $\hbar\omega$. A three-dimensional Weyl valley is charactrized by a universal value of circular photocurrent generation rate, and the coefficient $\beta$ in Eq.~(\ref{beta}) is given by \cite{Hasan2017,Grushin2018,Juan:2017tm,Golub-Weyl-2}
\begin{equation}
\label{universal}
\beta = \mathcal C \beta_0\:,\:  \beta_0 = \frac{\pi}{3} \frac{ e^3}{h^2}\:,
\end{equation}
where $h = 2 \pi \hbar$ and $\mathcal C=-4$ is topological charge of the node. \addMisha{Note that the negative values of ${\cal C}, \ell$ and $v_p$ correspond to the LR enantiomorphous pair with Rh atoms positioned in the left-handed type and Si in the right-handed type structure. For the RL pair, these values are positive.} In Fig.~\ref{fig:cpge}b the photocurrent generation rate is shown as a function of the light frequency. One can see that, at $\hbar\omega\to 0$, the rate tends to a value of $-4$ indicated by the thin line described by Eq.~\eqref{universal}. 

With increasing energy $\hbar \omega$ the ratio $\beta/\beta_0$ deviates noticeably from $-4$ because of the nonlinear terms in the energy dispersion. For $\beta$, the peculiarity near $\hbar \omega = 90$ meV is absent because the singularity in the density of states is compensated by the difference $v_{n'{\bm k};z} - v_{n{\bm k};z}$ in Eq.~(\ref{gamma:nn'}). Under steady-state excitation, the circular photocurrent is macroscopically described by
\begin{equation} \label{gamma}
{\bm j} = {\rm i}  [{\bm e} \times {\bm e}^*] \gamma E^2_0\:.
\end{equation}
The coefficient $\gamma$ is the sum $\sum_{n'n} \gamma_{n'n}$ where the partial contributions are given by Eq. (\ref{gamma:nn'}) where the difference of velocities is replaced by~(cf. Ref.~\cite{IvchenkoPikus,Ivchenko,nestoklon})
\[
v_{n'{\bm k};z} \tau_{n'} - v_{n{\bm k};z} \tau_n\:,
\]
where $\tau_n$ is the momentum relaxation time of electrons in the band $n$, in general it is energy-dependent. Since for different bands values of $\tau_n$ do not coincide the frequency dependence of the CPGE current (\ref{gamma}) acquires a feature similar to that in Fig.~\ref{fig:cpge}a.

The CPGE in RhSi was calculated by Congcong Le et al. \cite{Sun2020}. Figure 1d in this paper shows the ratio $\beta(\hbar \omega)/\beta_0$ as a function of $\hbar \omega$ in the interval 0.1 -- 2 eV. One can see that in Fig.~\ref{fig:cpge}b, in agreement with \cite{Sun2020}, this ratio decreases in the interval 0.1 -- 0.2 eV. In contrast to Ref. \cite{Hasan2017} the developed ${\bm k}\cdot {\bm p}$ model allowed us to perform the calculation in the low-frequency region $\hbar \omega \leq 0.1$ eV. Use of the isotropic approximation for the band structure significantly reduces numerical errors in evaluation of the CPGE injection rate and allows us to study the deviations from the quantized value~\eqref{universal}, cf. Ref.~\cite{Grushin2018}.

\emph{Results and conclusion.} We have developed a $6 \times 6$ effective Hamiltonian $\mathcal{H}(\bm{k})$ describing electron states near the $\Gamma$ point in chiral semimetals (e.g., RhSi, CoSi) with the space group P2$_1$3 and point group T. The Hamiltonian includes both linear in $\bm{k}$ Rashba terms and quadratic in the wave vector contributions, accounting for spin-orbit coupling and band degeneracies. The model is parameterized using tight-binding calculations based on previous works. 

The calculated band structure near the $\Gamma$-point reveals a quartet $\Gamma_6 + \Gamma_7$ and a doublet $\Gamma_5$ due to the spin-orbit coupling. We have computed the absorption spectrum, which at low frequencies is a linear function of the frequency $\omega$, characteristic of bulk systems with $\bm{k}$-linear dispersion. The circular photogalvanic effect, the generation of a photocurrent sensitive to the light circular polarization, is calculated and analyzed within the developed $\bm k\cdot \bm p$ Hamiltonian approach. The photocurrent generation rate approaches a universal quantized value at low frequencies. The $\bm{k}$-linear Rashba terms are essential for the CPGE, enabling quantized photocurrents in the low-frequency limit. 

\emph{Acknowledgement.} This work was supported by the RSF grant 23-12-00142. The authors are grateful to L.E. Golub for valuable discussions.

\end{document}


\title{Online Supplementary Materials to \\
Photocurrents induced by ${\bm k}$-linear terms in semiconductors and semimetals}
\author{M.~M.~Glazov, E.~L.~Ivchenko}
\affiliation{Ioffe Institute, 194021, St.-Petersburg, Russia}

  \maketitle
  
  \tableofcontents

\section{$6\times6$ Matrices $\mathcal J_i$} \label{App1}

For brevity, we present not three separate matrices ${\cal J}_i$, but one matrix:
\begin{equation} \label{kJ}
{\bm k}\cdot \bm{\mathcal{J}}=   \left[ \begin{array}{cccccc} \frac32 k_z & \frac{\sqrt{3}}{2} k_- & 0 & 0 &  -\frac12 \sqrt{\frac32}k_- &0 \\ \frac{\sqrt{3}}{2} k_+& \frac12 k_z & k_- & 0 &  \frac{ k_z}{\sqrt{2}} & -\frac{k_-}{2\sqrt{2}} \\ 0 & k_+ & - \frac12 k_z & \frac{\sqrt{3}}{2} k_- & \frac{k_+}{2\sqrt{2}} & \frac{k_z}{\sqrt{2}}    \\ 0 & 0 & \frac{\sqrt{3}}{2} k_+ & - \frac32 k_z & 0 & \frac12 \sqrt{\frac32} k_+ \\ \sqrt{\frac32}k_+&  \frac{1}{\sqrt{2}} k_z &-\sqrt{2} k_- &0&k_z& k_-\\ 0 & \sqrt{2}k_+ &\frac{1}{\sqrt{2}} k_z &\sqrt{\frac32} k_-&k_+& - k_z
\end{array} \right] 
\end{equation}
Clearly, the matrix ${\cal J}_i$ is obtained from matrix (\ref{kJ}) by setting $k_i = 1$ and $k_{i'} = 0$ for $i'\neq i$. Note that in deriving Eq.~(4) of the main text it was assumed that splitting $\Delta$ is much smaller than energy distance to other bands. According to numerical calculations \cite{Grushin2018,Sun2020} the parameter $\Delta$ is positive.

In Ref.~\cite{Grushin2018}, Eq.~(4), a contribution of linear in ${\bm k}$ terms to $4\times4$ Hamiltonian for $\Gamma_6 + \Gamma_7$ band is given. Here the basis $3/2, - 1/2, - 3/2, 1/2$ is used. Comparing the matrix (4) in \cite{Grushin2018} with the $4 \times 4$ block composed of elements in the first four columns and rows of matrix (\ref{kJ}), we get that coefficient $\hbar v_0$ in Eq.~(5) of the main text and coefficients $a,b$ in Ref.~\cite{Grushin2018} are related by $a = - 3b = \frac32 \hbar v_0$. The block of first four columns and rows with quadratic terms represents the well-known Luttinger Hamiltonian for semiconductors of the cubic symmetry T$_d$ (GaAs, InSb, ZnSe...) and O$_h$ (Ge, Si, diamond). Electronic states in $\Gamma_6 + \Gamma_7$ and $\Gamma_5$ bands can be considered independently only if $\hbar |v_0| k, |B|k^2, |D|k^2 \ll {\Delta}$. 

\section{General Form of Linear in the Wavevector Terms} \label{App2}
Using the method of invariants, we derive a general expression for terms linear in the wave vector  in the $\left(1,\frac12\right)$-fermion Hamiltonian in the T symmetry crystals. For this we represent matrix ${\cal H}({\bm k})$ as four blocks of dimensions $4\times4, 4\times2, 2\times4$ and $2\times2$:
\begin{equation} \label{Hgeneral}
{\cal H}({\bm k})  = \left[ \begin{array}{cc}  {\cal H}_{4 \times 4}({\bm k})  & {\cal H}_{4 \times 2}({\bm k})  \\  {\cal H}_{2 \times 4}({\bm k}) & {\cal H}_{2 \times 2}({\bm k}) \end{array} \right]\:.
\end{equation}
Each block can be expanded in basis matrices transforming according to representation $\Gamma_6 + \Gamma_7$ and odd under time reversal. There are nine such matrices in $4\times4$ block, six in $4\times2$ block and three in $2\times2$ block:
\begin{eqnarray}
&&{\cal H}^{(1)}_{4\times 4}({\bm k})  = \hbar \sum\limits_{l=1}^3 v_l^{(4\times4)} \hspace{0.5 mm} {\bm k} \cdot  {\bm J}_{4\times4}^{(l)}\:, \\
&&{\cal H}^{(1)}_{4\times 2}({\bm k})  = {\cal H}^{(1)\dag}_{2\times 4}({\bm k})  = \hbar \sum\limits_{l=1}^2 v_l^{(4\times2)} \hspace{0.5 mm} {\bm k} \cdot  {\bm J}_{4\times2}^{(l)}\:,\: {\cal H}^{(1)}_{2\times 2}({\bm k})  = \hbar v^{(2\times 2)} {\bm k}\cdot{\bm \sigma}\:,
\end{eqnarray}
where $v^{(n' \times n)}_l$ are constant coefficients with dimension of velocity.
The basis matrices $J^{(l)}_{4\times4;i}$ are expressed through the angular momentum matrices $J_i$ with $J=3/2$ \cite{Ivchenko}:
\begin{equation}
J^{(1)}_{4 \times 4;i} = J_i\:,\:J^{(2)}_{4 \times 4} = J_i^3\:,\: J^{(3)}_{4 \times 4{;i}} = V_i \equiv \frac12 [ J_i (J_{i+1}^2 - J^2_{i+2}) + (J_{i+1}^2 - J^2_{i+2})  J_i  ]\:.
\end{equation}
The six matrices $J_{4\times2;i}^{(l)}$ are presented, similarly to Eq.~(\ref{kJ}), similarly to Eq. (1), in the compact form:
\begin{equation}
{\bm k} \cdot {\bm J}^{(1)}_{4 \times 2} =  {\bm k} \cdot {\bm J}^{(1)\dag}_{2 \times 4}  = \left[ \begin{array}{cc} - \frac{k_-}{\sqrt{2}} &0   \\ \sqrt{\frac23} k_z & -\frac{k_-}{\sqrt{6}} \\  \frac{k_+}{\sqrt{6}}  &  \sqrt{\frac23} k_z   \\ 0 & \frac{k_+}{\sqrt{2}} \:, \end{array} \right] ,\:  
{\bm k} \cdot {\bm J}^{(2)}_{4 \times 2} =  \left[ \begin{array}{cc}  - \sqrt{\frac32}k_-& 0 \\  \frac{1}{\sqrt{2}}k_z& \sqrt{2}k_- \\ - \sqrt{2}k_+ & \frac{1}{\sqrt{2}}k_z \\ 0 & \sqrt{\frac32}k_+\end{array} \right]\:, \nonumber
\end{equation}
One can check that $4\times2$ blocks in (\ref{kJ}) and matrix ${\bm k} \cdot {\bm J}^{(1)}_{4 \times 2}$ differ only by a factor of $\sqrt{3}/2$.

\section{Tight-binding model of Ref.~\cite{Hasan2017}}\label{App3}

We use the tight-binding model presented in Sec. C of the Supplementary Materials to Ref.~\cite{Hasan2017}. The tight-binding Hamiltonian in the N=$198$ space group, $\mathcal{H}_{198}(\bm k)$, can be expressed as a sum of the spin-independent part $\tilde{\mathcal{H}}_{198}(\bm k)$, the spin-orbit terms $V_{r,i}(\bm k), V_{s,i}(\bm k)$ ($i=1,2,3$), and the spin-independent second-nearest-neighbor hopping:
\begin{equation}
\mathcal{H}_{198}(\bm k) = \tilde{\mathcal{H}}(\bm k)_{198} + \sum_{i=1,2,3}V_{r,i}(\bm k) + V_{s,i}(\bm k)
\label{eq:tb}
\end{equation}
where
\begin{eqnarray}
\tilde{\mathcal{H}}_{198}(\bm k) &=& v_{1}\bigg[\tau^{x}\cos\left(\frac{k_{x}}{2}\right)\cos\left(\frac{k_{y}}{2}\right) + \tau^{x}\mu^{x}\cos\left(\frac{k_{y}}{2}\right)\cos\left(\frac{k_{z}}{2}\right) + \mu^{x}\cos\left(\frac{k_{z}}{2}\right)\cos\left(\frac{k_{x}}{2}\right)\bigg] \nonumber \\
&+&  v_{p}\bigg[\tau^{y}\mu^{z}\cos\left(\frac{k_{x}}{2}\right)\sin\left(\frac{k_{y}}{2}\right) + \tau^{y}\mu^{x}\cos\left(\frac{k_{y}}{2}\right)\sin\left(\frac{k_{z}}{2}\right) + \mu^{y}\cos\left(\frac{k_{z}}{2}\right)\sin\left(\frac{k_{x}}{2}\right)\bigg], 
\end{eqnarray}
\begin{eqnarray}
V_{r1}(\bm k)&=&v_{r1}\bigg[\tau^{y}\mu^{z}\sigma^{y}\cos\left(\frac{k_{x}}{2}\right)\cos\left(\frac{k_{y}}{2}\right) + \tau^{y}\mu^{x}\sigma^{z}\cos\left(\frac{k_{y}}{2}\right)\cos\left(\frac{k_{z}}{2}\right) + \mu^{y}\sigma^{x}\cos\left(\frac{k_{z}}{2}\right)\cos\left(\frac{k_{x}}{2}\right)\bigg]  \\
V_{r2}(\bm k)&=&v_{r2}\bigg[\tau^{y}\sigma^{z}\cos\left(\frac{k_{x}}{2}\right)\cos\left(\frac{k_{y}}{2}\right) + \tau^{x}\mu^{y}\sigma^{x}\cos\left(\frac{k_{y}}{2}\right)\cos\left(\frac{k_{z}}{2}\right) + \tau^{z}\mu^{y}\sigma^{y}\cos\left(\frac{k_{z}}{2}\right)\cos\left(\frac{k_{x}}{2}\right)\bigg] \nonumber \\
V_{r3}(\bm k)&=&v_{r3}\bigg[\tau^{y}\mu^{z}\sigma^{x}\sin\left(\frac{k_{x}}{2}\right)\sin\left(\frac{k_{y}}{2}\right) + \tau^{y}\mu^{x}\sigma^{y}\sin\left(\frac{k_{y}}{2}\right)\sin\left(\frac{k_{z}}{2}\right) + \mu^{y}\sigma^{z}\sin\left(\frac{k_{z}}{2}\right)\sin\left(\frac{k_{x}}{2}\right)\bigg] \nonumber \\
V_{s1}(\bm k)&=&v_{s1}\bigg[\tau^{x}\sigma^{x}\sin\left(\frac{k_{x}}{2}\right)\cos\left(\frac{k_{y}}{2}\right) + \tau^{x}\mu^{x}\sigma^{y}\sin\left(\frac{k_{y}}{2}\right)\cos\left(\frac{k_{z}}{2}\right) + \mu^{x}\sigma^{z}\sin\left(\frac{k_{z}}{2}\right)\cos\left(\frac{k_{x}}{2}\right)\bigg] \nonumber \\
V_{s2}(\bm k)&=&v_{s2}\bigg[\tau^{x}\sigma^{y}\cos\left(\frac{k_{x}}{2}\right)\sin\left(\frac{k_{y}}{2}\right) + \tau^{x}\mu^{x}\sigma^{z}\cos\left(\frac{k_{y}}{2}\right)\sin\left(\frac{k_{z}}{2}\right) + \mu^{x}\sigma^{x}\cos\left(\frac{k_{z}}{2}\right)\sin\left(\frac{k_{x}}{2}\right)\bigg] \nonumber \\
V_{s3}(\bm k)&=& v_{s3}\bigg[\tau^{x}\mu^{z}\sigma^{z}\cos\left(\frac{k_{x}}{2}\right)\sin\left(\frac{k_{y}}{2}\right) - \tau^{y}\mu^{y}\sigma^{x}\cos\left(\frac{k_{y}}{2}\right)\sin\left(\frac{k_{z}}{2}\right) + \tau^{z}\mu^{x}\sigma^{y}\cos\left(\frac{k_{z}}{2}\right)\sin\left(\frac{k_{x}}{2}\right)\bigg],  \nonumber  
\end{eqnarray}
and
\begin{equation}
V_{2}(\bm k)=v_{2} {[\cos\left(k_{x}\right) + \cos\left(k_{y}\right) + \cos\left(k_{z}\right)]\hat{1}}.
\end{equation}
Here we use the notations of Ref.~\cite{Hasan2017} with $x$, $y$ and $z$ being the cubic axes, $k_i$ being the components of the wavevector $\bm k$ in the units of the lattice constant $a_0$,  and $\tau^{i}\mu^{j}\sigma^{k}$ standing for the Kronecker product of the three spin-Pauli matrixes ($i,j,k$ are $x$, $y$, $z$), $\hat{1}$ is the $8\times 8$ identity matrix.

Table~I the main text shows the parametrization of the $\bm k\cdot \bm p$ model via the tight-binding parameters. The spin-orbit splitting parameter $\Delta$ can be presented as
\begin{equation}
\label{Delta}
\Delta = \left|v_1 + 3 v_2 - v_{r1} + v_{r2} - 
 2 \sqrt{v_1^2 + v_1 v_{r1} + v_{r1}^2 - v_1 v_{r2} + v_{r1} v_{r2} + v_{r2}^2}\right|.
\end{equation}
\section{Energy and Eigenstates in the Isotropic Approximation}\label{App4}
For $k_z \parallel [001]$, the matrix (5) of the main text splits into blocks of dimension 1 and 2. This allows one writing the eigenenergies in the analytical form
\begin{eqnarray}
&& E_{3/2}(k) = \frac32 a k_z + (A-B) k_z^2,\: E_{-3/2} = - \frac32 a k_z + (A-B) k_z^2\:,
\end{eqnarray}
\begin{eqnarray}&&
E_{1/2;1}(k) = \frac{u_{11} + u_{22}}{2} + \sqrt{\left( \frac{u_{11} - u_{22}}{2} \right)^2 + u_{12}^2}\:, \\
&&E_{1/2;2}(k) = \frac{u_{11} + u_{22}}{2} - \sqrt{\left( \frac{u_{11} - u_{22}}{2} \right)^2 + u_{12}^2}\:, \nonumber \\ &&
E_{-1/2;1}(k) = \frac{v_{11} + v_{22}}{2} + \sqrt{\left( \frac{v_{11} - v_{22}}{2} \right)^2 + v_{12}^2}\:, \nonumber \\ &&
E_{-1/2;1} (k)= \frac{v_{11} + v_{22}}{2} - \sqrt{\left( \frac{v_{11} - v_{22}}{2} \right)^2 + v_{12}^2} \:, \nonumber 
\end{eqnarray}
where
\begin{eqnarray} \label{uv}
&& u_{11} = \frac12 a k_z + (A + B) k_z^2\:,\: u_{22} = - \Delta + a k_z + A  k_z^2\:,\: u_{12} = \frac{1}{\sqrt{2}} a k_z - \sqrt{2} B k_z^2\:,\\ &&v_{11} = -\frac12 a k_z + (A + B) k_z^2\:,\: v_{22} =- \Delta  - a k_z + A  k_z^2\:,\: v_{12} =  \frac{1}{\sqrt{2}} a k_z + \sqrt{2} B k_z^2 \:. \nonumber 
\end{eqnarray}
The eigencolumns of the operator $\mathcal{H}(\bm{k} \parallel z)$ in the basis $\Psi_j$ ($j=1-6$) have the form:
\begin{equation} \label{CC}
\hat{C}_{3/2}({\bm k}) = \left[ \begin{array}{c} 1 \\ 0 \\ 0\\ 0\\ 0\\ 0 \end{array} \right], \hat{C}^{(\pm)}_{1/2}({\bm k}) = \left[ \begin{array}{c} 0 \\ C_{1/2;4}^{(\pm)} \\ 0\\ 0\\ C_{1/2;2}^{(\pm)}\\ 0 \end{array} \right], \hat{C}^{(\pm)}_{-1/2} ({\bm k})= \left[ \begin{array}{c} 0 \\0 \\ C_{-1/2;4}^{(\pm)} \\ 0\\ 0\\ C_{-1/2;2}^{(\pm)} \end{array} \right],  \hat{C}_{-3/2}({\bm k}) = \left[ \begin{array}{c} 0 \\ 0 \\ 0\\ 0\\ 0\\ 1 \end{array} \right]\:. 
 \end{equation}
Here
\begin{eqnarray}
&&  C_{1/2;4}^{(+)} = \sqrt{\frac{1 + U}{2}}\:,\: C_{1/2;2}^{(+)} = \frac{u_{12}}{|u_{12}|} \sqrt{\frac{1 - U}{2}}\:, \\
&&  C_{1/2;4}^{(-)} = - \frac{u_{12}}{|u_{12}|} \sqrt{\frac{1 - U}{2}}\:, \:   C_{1/2;2}^{(+)} = \sqrt{\frac{1 + U}{2}}\:, \nonumber \\
&&  C_{-1/2;4}^{(+)} = \sqrt{\frac{1 + V}{2}}\:,\: C_{-1/2;2}^{(+)} = \frac{v_{12}}{|v_{12}|} \sqrt{\frac{1-V}{2}}\:,\nonumber \\
&&  C_{-1/2;4}^{(-)} = - \frac{v_{12}}{|v_{12}|} \sqrt{\frac{1-V}{2}}\:, \:   C_{-1/2;2}^{(+)} = \sqrt{\frac{1+V}{2}}\:, \nonumber
\end{eqnarray}
\begin{eqnarray}
U = \frac{u_-}{\sqrt{u_-^2+|u_{12}|^2}}\:,\: V = \frac{v_-}{\sqrt{v_-^2+|v^2_{12}|^2}}\:,
\end{eqnarray}
$u_- = (u_{11} - u_{22})/2, v_- = (v_{11} - v_{22})/2$.

\begin{figure}[h]
\includegraphics[width=\textwidth]{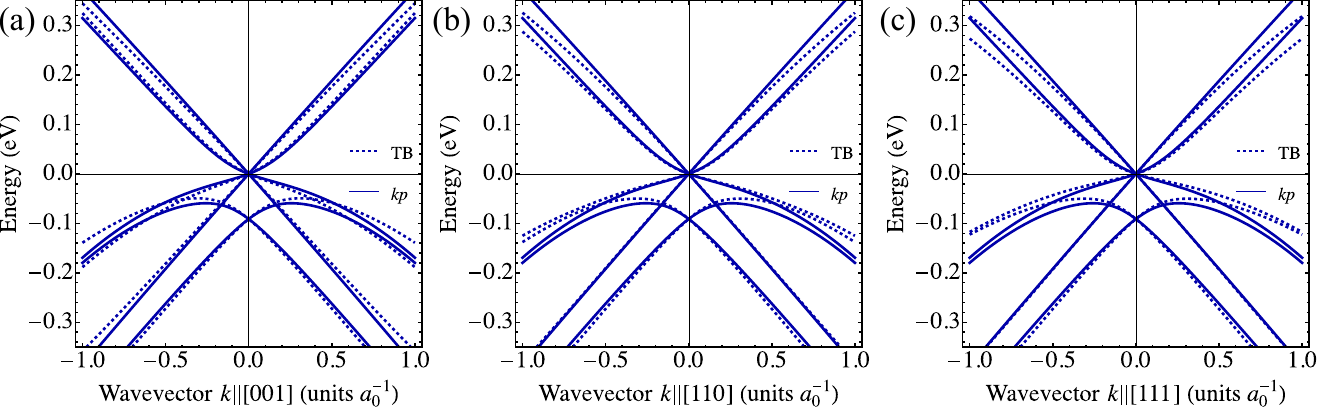}
\caption{(a-c) Spectrum near the $\Gamma$ point calculated with spin-orbit interaction in the tight-binding method (dashed) and in the $\bm{k} \cdot \bm{p}$ model (solid curves) for three different directions of the quasi-wave vector. The parameters of the $\bm{k} \cdot \bm{p}$ model obtained by fitting the tight-binding calculation are given in Table~{I of the main text}.}\label{fig:spec}
\end{figure}

In the isotropic approximation, when $D = \sqrt{3} B$, the energy dispersion $E_{\pm 3/2}$, $E_{\pm 1/2, 1}$, $E_{\pm 1/2, 2}$ is obtained by replacing $k_z$ with $k = |\bm{k}|$. Figures~\ref{fig:spec}(a-c) show a comparison of the $\bm{k} \cdot \bm{p}$ calculation (solid lines) with the tight-binding calculation for three non-equivalent directions $\bm{k} \parallel [001]$ (a), $\bm{k} \parallel [110]$ (b), and $\bm{k} \parallel [111]$ (c) demonstrating good agreement between the $\bm k\cdot\bm p$ model and the tight-binding approximation and applicability of the isotropic approximation for the energy spectrum.

\section{Absorption and CPGE in the Isotropic Approximation}\label{App5}

The $6\times6$ velocity operator 
\begin{equation}
\hat{\bm v}({\bm k}) = \frac{1}{\hbar} \frac{\partial \mathcal{H}({\bm k})}{\partial {\bm k}} \nonumber 
\end{equation}
can be conveniently presented as a sum $\hat{\bm v}^{(1)}({\bm k}) + \hat{\bm v}^{(2)}({\bm k})$ where 
\begin{eqnarray}
&& {\bm e}\cdot \hat{\bm v}^{(1)}= \frac{1}{\hbar} \mathcal{H}^{(1)}({\bm e}), \nonumber\\ &&
{\bm e} \cdot \hat{\bm v}^{(2)}({\bm k}) = \frac{1}{\hbar} \frac{ \partial \mathcal{H}^{(2)} }{\partial k_{\eta}\partial k_{\zeta}} \left( k_{\eta} e_{\zeta} + k_{\zeta}e_{\eta}  \right), 
\end{eqnarray}
and $\mathcal{H}^{(1)}({\bm k}), \mathcal{H}^{(2)}({\bm k})$ are the contributions to the effective Hamiltonian, linear and quadratic in $\bm{k}$, respectively.

Analytical calculations are significantly simplified in the isotropic approximation. Let us introduce a coordinate system $x',y',z'$ rotated relative to $x,y,z$ by the axial angle $\varphi$ and polar angle $\theta$ of the vector $\bm{k}$, so that $z' \parallel \bm{k}$. In the isotropic approximation, the identity holds:
\begin{equation}\label{ev}
| {\bm e}\cdot {\bm v}_{mn}({\bm k})|^2 = | {\bm e}'\cdot {\bm v}_{mn}({\bm k}')|^2,
\end{equation}
where ${\bm e}'$ is the light polarization vector with components $e_{x'}, e_{y'}, e_{z'}$ and ${\bm k}'$ is the vector $(0,0,k)$. Therefore, the average of the left hand side of (\ref{ev}) over all directions of $\bm{k}$ is the average of the right hand side of (\ref{ev}) over the angles $\varphi$ and $\theta$, on which the vector ${\bm e}'$ depends. Here, the following simple relations hold
\[
\overline{|e'_{+}|^2} = {\overline{|e'_{-}|^2} }=  2 \overline{|e'_z|^2} = \frac{2}{3}\:{.}
\]
Table~\ref{tab:select} shows the polarization dependence of the matrix elements of the velocity operator ${\bm e}' \cdot {\bm v}_{mn}({\bm k}')$ for $ {\bm k}' = (0,0,k)$. As an example, we give the expression for one of the factors $v_{n',n}$:
\begin{equation} 
v_{1/2,3/2} = \hat{C}^{(+)\dag}_{1/2} \hat{\bm v}(k) \hat{C}_{3/2} = \frac{1}{\hbar} \left[ \frac{\sqrt{3}}{2} a C^{(+)}_{1/2;4} + \left( \sqrt{\frac32} a + \frac{Dk}{\sqrt{2}}\right) C^{(+)}_{1/2;2}\right]\ .
\end{equation}

\begin{table}
\caption{Selection rules for the matrix elements of the velocity operator calculated between pairs of states $|\pm 1/2, +\rangle$ and the quartet of states $|3/2\rangle; |\pm 1/2,-\rangle; |- 3/2\rangle$. Here $e'_{{\pm}} = e'_x \pm e'_y$, factors $v_{n'n}$ are real functions of $k$.}\label{tab:select}
 \begin{center}
\begin{tabular}{|l||l|l|l|l||} \hline 
& $3/2$ & $1/2;-$ & $-1/2;-$ &$-3/2$ \\ \hline \hline
$1/2;+$&$v_{1/2;3/2}\ e'_+$&$v_{1/2;1/2}\ e'_z$&$v_{1/2;-1/2}\ e'_-$&0\\ \hline
$-1/2;+$&0&$v_{-1/2;1/2}\ e'_+$&$v_{-1/2;-1/2}\ e'_z$&$v_{-1/2;-3/2}\ e'_-$ \\ \hline
\end{tabular}
\end{center}
\end{table}

Accordingly, the contribution to the absorption coefficient related to transitions $n \to n'$, Eq.~(9) in the main text, can be written as
\begin{subequations}
\begin{eqnarray} 
K_{n'n} &=& \frac{4\pi^2e^2}{\omega c n_{\omega}V}\sum\limits_{\bm k} | {\bm e}\cdot {\bm v}_{n'n}({\bm k})|^2 \left( f_{n k} - f_{n'k}\right) \delta \left[ E_{n'}(k) - E_n(k)- \hbar \omega \right]\label{Knum}
\\ &=& \frac{4\pi^2e^2}{\omega c n_{\omega}} \left[ \overline{| {\bm e}\cdot {\bm v}_{n'n}({\bm k})|^2} \left( f_{n k_{n'n}} - f_{n' k_{n'n}}\right) \right]_{k = k_{n'n}} g_{n'n}(\hbar \omega){,}
\end{eqnarray}
\end{subequations}
where the wave vector $k_{n'n}$ satisfies the equation $E_{n'}(k_{n'n}) - E_n(k_{n'n}) = \hbar \omega$, and the reduced density of states is:
\[
g_{n'n}(\hbar \omega) = \frac{1}{2 \pi^2} \left(\frac{k^2}{|E'_m - E'_n|}\right)_{k = k_{n'n}}\:.
\]

The contribution to the photocurrent quadratic in the field amplitude $E_0$ inside the crystal and sensitive to circular light polarization is written for the point group T as [see Eq. (13) of the main text]:
\begin{equation}
{\bm j}_{\rm CPGE}= \gamma {\rm i} [{\bm e} \times {\bm e}^*] E^2_0\:.
\end{equation}
The partial coefficient $\gamma_{n'n}$ for the direct transitions between the valence band $n$  and the the conduction band $n'$ has the form
\begin{equation}
\label{gamma:nn'}
\gamma_{n'n} = \frac{ 2\pi} {\hbar} \frac{e^3}{\omega^2}\sum\limits_{\bm k} { (v_{n'{\bm k};z} \tau_{n'} - v_{n{\bm k};z} \tau_n)} |{\bm e}_{\sigma_+}\cdot {\bm v}_{n'n}({\bm k})|^2 \left( f_{n \bm k} - f_{n'\bm k}\right) \delta \left[ E_{n'}(\bm k) - E_n(\bm k)- \hbar \omega \right]\:.
\end{equation}
Note that in the isotropic approximation 
\[
{\bm v}_{n{\bm k}} = \frac{1}{\hbar} \frac{\partial E_n(\bm k)}{\partial {\bm k}} = \frac{1}{\hbar} \frac{\partial E_n(k)}{\partial k} \frac{\bm k}{k}\:.
\]

The expression for $\gamma_{n'n}$ can be simplified in the isotropic energy spectrum approximation by choosing $\bm k'\parallel z$ and making a transition to the corresponding axes:
\begin{equation}
\gamma_{n'n} = \frac{ 2\pi} {\hbar} \frac{e^3}{\omega^2}\sum\limits_{\bm k}  (v_{n'{\bm k};z} \tau_{n'} - v_{n{\bm k};z} \tau_n) |{\bm e}'_{\sigma_+}\cdot {\bm v}_{n'n}({\bm k}')|^2 \left( f_{n k} - f_{n'k}\right) \delta \left[ E_{n'}(k) - E_n(k)- \hbar \omega \right]\:,
\end{equation}
where ${\bm e}'_{\sigma_+}$ is the vector with the components
\[
\frac{{\rm e}^{{\rm i} \varphi}}{\sqrt{2}} ( \cos{\theta}, {\rm i}, \sin{\theta})\:,
\]
onto the axes $x',y',z'$. Note that,
\begin{equation}
 | e'_{\sigma_+; x'} \pm {\rm i} e_{\sigma_+; y'}  |^2 = \frac{(1 \mp \cos{\theta})^2}{2} \:,\: |e'_{\sigma_+; z'}|^2 = \frac{\sin^2{\theta}}{2}\:, \nonumber
\end{equation}
and 
\[
\overline{k_z  | e_{\sigma_+; x'} \pm {\rm i} e_{\sigma_+; y'}  |^2} = \mp \frac{k}{3}\:,\: \overline{k_z  |e_{\sigma_+; z'}|^2} = 0\:,
\]
where the overline means averaging over $\theta$ and $\varphi$.

\begin{figure}[h]
\includegraphics[width=0.5\linewidth]{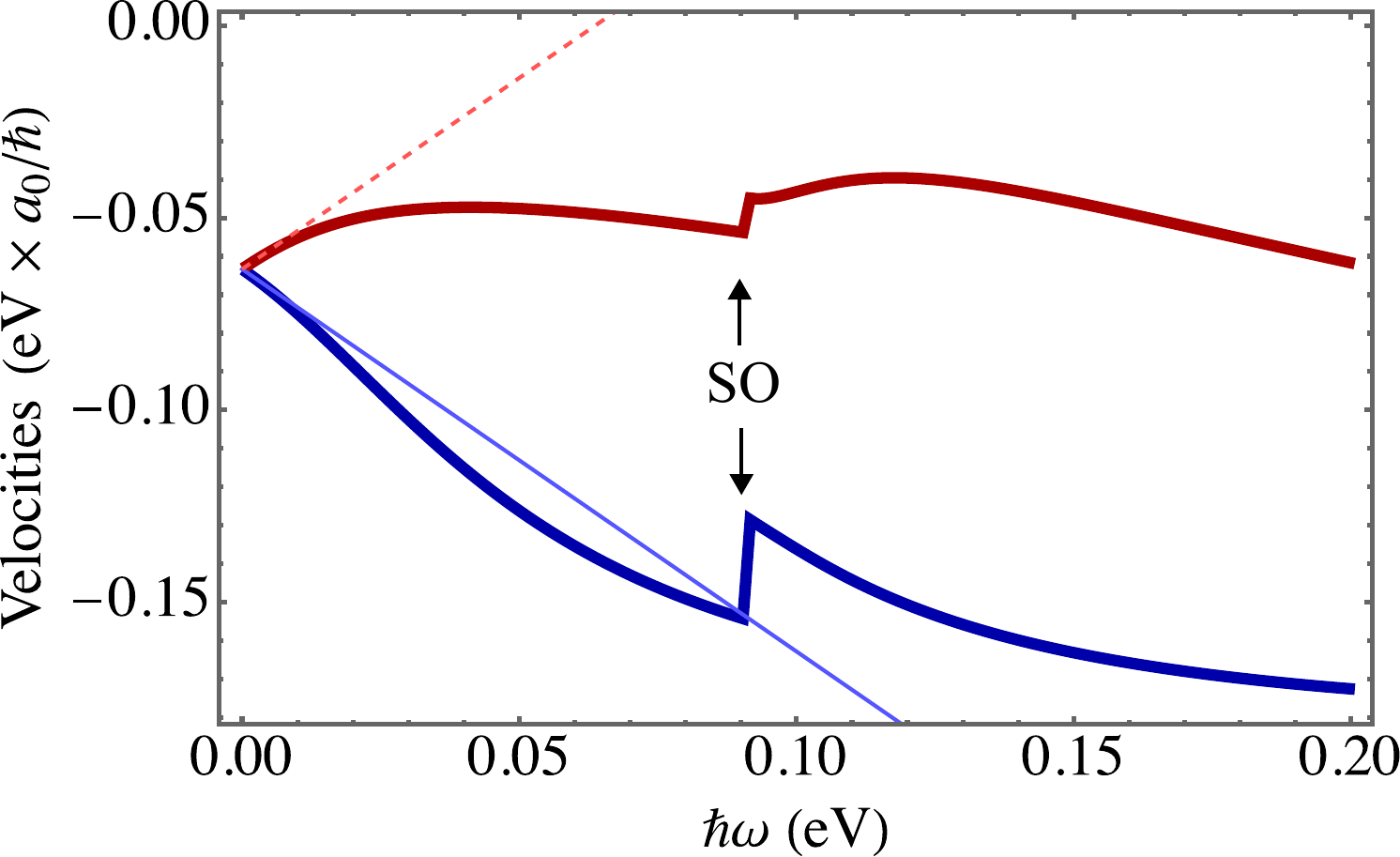}
\caption{Velocities $\bar v_c$ (thick blue curves), $\bar v_v$ (thick red curves) calculated using formulas~\eqref{vcvv} for interband optical transitions. Thin lines show asymptotics at $\hbar\omega \to 0$, Eqs.~\eqref{vcvv:SO}.
Parameters of the $\bm{k} \cdot \bm{p}$ model used in calculations are given in Table~\ref{tab:kp}. }\label{fig:cpge}
\end{figure}

Figure~\ref{fig:cpge} presents the results of calculating the circular photocurrent excitation spectrum versus the photon energy $\hbar\omega$. The plots show quantities $\bar v_c$ (blue curves), $\bar v_v$ (red curves), introduced according to:
\begin{subequations}
\label{vcvv}
\begin{align}
\bar v_c &= \mathcal A^{-1} \sum\limits_{\bm k,n,n'}  {v}_{n'{\bm k}; z} |{\bm e}_{\sigma_+}\cdot {\bm v}_{n'n}({\bm k})|^2 \left( f_{n \bm k} - f_{n'\bm k}\right) \delta \left[ E_{n'}(\bm k) - E_n(\bm k)- \hbar \omega \right],\\
\bar v_v &= -\mathcal A^{-1} \sum\limits_{\bm k,n,n'}  {v}_{n{\bm k}; z} |{\bm e}_{\sigma_+}\cdot {\bm v}_{n'n}({\bm k})|^2 \left( f_{n \bm k} - f_{n'\bm k}\right) \delta \left[ E_{n'}(\bm k) - E_n(\bm k)- \hbar \omega \right],
\end{align}
where 
\[
\mathcal A = \left(\frac{4\pi^2e^2}{\omega c n_{\omega}}\right)^{-1}K.
\]
The physical meaning of $\bar v_c$ ($\bar v_v$) is the charge carrier velocity in the conduction (valence) band acquired per absorbed photon~\cite{nestoklon}.
\end{subequations}
The thin straight lines in Fig.~\ref{fig:cpge} show the results of analytical calculation of the leading contributions to photocurrent at $\hbar\omega \to 0$, caused by presence of linear and quadratic terms in energy spectrum: 
\begin{equation}
\label{vcvv:SO}
\bar{\bm v}_c= \left[\frac{\ell}{6\hbar} + \xi \hbar\omega\right]\mathrm i [\bm e\times \bm e^*], \quad 
\bar{\bm v}_v= \left[\frac{\ell}{6\hbar} - \xi \hbar\omega\right]\mathrm i [\bm e\times \bm e^*],
\end{equation}
where
\begin{equation}
\xi = \frac{\ell}{3\hbar \Delta} + \frac{L+M/2}{\hbar\ell}.
\end{equation}
These dependencies are approximately satisfied for not too high frequencies. Deviations from these analytical expressions occur in the same frequency range where the photocurrent generation rate deviates from the universal value given by Eq. (12) of the main text, see also Fig. 2(b) of the main text.